\documentclass[sn-mathphys,Numbered]{sn-jnl}

\usepackage{graphicx}%
\usepackage{multirow}%
\usepackage{amsmath,amssymb,amsfonts}%
\usepackage{amsthm}%
\usepackage{mathrsfs}%
\usepackage[title]{appendix}%
\usepackage{xcolor}%
\usepackage{textcomp}%
\usepackage{manyfoot}%
\usepackage{booktabs}%
\usepackage{algorithm}%

\usepackage{algorithmicx}%
\usepackage{algpseudocode}%
\usepackage{listings}%
\usepackage{cancel}
\usepackage{hyperref}
\usepackage{simpler-wick} 
\usepackage{accents}
\usepackage{xcolor}
\usepackage{appendix}
\usepackage{multirow}
\usepackage{physics} 
\usepackage{lipsum}
\usepackage{bm}
\usepackage{upgreek}
\usepackage{tikz}
\usepackage{multirow}
\usepackage{subfigure}
\usetikzlibrary{decorations.pathmorphing, patterns,shapes}

\begin{document}

\title[Article Title]{Relations between closed string amplitudes and mixed string amplitudes at tree-level}


\author{\fnm{Aphiwat} \sur{Yuenyong}}\email{aphiwaty64@nu.ac.th}

\author*{\fnm{Pongwit} \sur{Srisangyingcharoen}}\email{pongwits@nu.ac.th}

\affil{\orgdiv{The Institute for Fundamental Study}, \orgname{Naresuan University}, \orgaddress{ \city{Phitsanulok}, \postcode{65000}, \country{Thailand}}}


\abstract{This paper investigates the relationships between closed and mixed string amplitudes at the tree level in string theory. Through the analytic continuation of complex variables, we establish a factorization of closed string amplitudes into those involving ($n-2$) open strings and a single closed string. Expressions for four-, five-, and six-point amplitudes are provided, along with systematic formulations for $n$ strings. The paper addresses possible correction terms arising from integration along infinite tubes during the Wick rotation process. In the field theory limit, the correction terms become negligible due to being of subleading order.}




\maketitle

\section{Introduction}\label{sec1}
String theory is often considered a candidate for the theory that describes all known interactions including gravity. Despite being debatable due to its peculiar features such as extradimensions and supersymmetries, the study of string theory is still useful, particularly in the area of scattering amplitudes. It has been established that in the limit of low energies, amplitudes in string theory reproduce those in QFT such as Yang-Mills \cite{Neveu:1971mu} and Einstein theory \cite{Scherk:1974ca,Yoneya:1974jg} plus $\alpha'$ corrections \cite{TSEYTLIN1986391, Koerber:2001uu, Metsaev:1986yb}. Understanding structures of string amplitudes does not only deepens our understanding of string theory but also contributes to advancements in field theories. A great example is a striking discovery made by Kawai, Lewellen and Tyle (KLT) which provides a relation between closed and open string amplitudes at tree level \cite{kawai1986relation}. Also, see \cite{Sondergaard:2011iv} for a review. The relation relates closed string amplitudes in terms of products of two open string amplitudes which provide alternative descriptions of gravity as the square of gauge theory in the low energy limit \cite{Bern:2002kj, Bern:2008qj, Bern:2010ue, Bern:2010yg}.

Another interesting structure is known as monodromy relation which relates partial open string amplitudes all together. This allow us to reduce the number of independent color-ordered amplitudes from $(n-1)!$ down to $(n-3)!$. A chosen set of $(n-3)!$ open string amplitudes are called minimal basis where the remaining amplitudes can be written in terms of this set of basis \cite{bjerrum2009minimal}. In field theory limit, $\alpha'$ $\to$ 0, the monodromy relations reduce to the BCJ relations of Bern, Carrasco and Johansson \cite{Bern:2008qj} and the Kleiss-Kuijf relation\cite{Kleiss:1988ne}. The monodromy relations among partial open string amplitudes can be captured by polygons in the complex plane \cite{Srisangyingcharoen:2020lhx}.

String amplitudes are described by the correlation functions evaluated on the Riemann surfaces on which the string states are introduced as local operators called vertex operators. At tree level, closed string states are inserted on a sphere which can be conformally mapped to the full complex plane while open string states are placed along the boundary of a disk or the upper half-plane. In this paper, we seek to find connections between closed string amplitudes and the amplitudes that involve scattering between both closed and open strings known as mixed string amplitudes. The general expression of $n$-point closed string amplitudes at tree-level are given as the integral \cite{Green:2012oqa}
    \begin{align}
        \mathcal{A}_n^{\text{cl}} = C_{S^2}(2\pi)^{D}\delta^{D}(\sum_i k_i)\int \frac{\abs{z_{ab}z_{ac}z_{bc}}^2}{dz_adz_bdz_c} \prod_{i=1}^{n}d^2z_i \prod_{1\leq j< l \leq n} \abs{z_j-z_l}^{\alpha'k_j \cdot k_l} F_n(z_i,\bar{z}_i), \label{e1}
    \end{align}
    where $C_{S^2}$ is a normalization constant and $z_{ij}$ = $z_i-z_j$. The function $F_n$ is a branch-free function that contains polarization and kinematic factors of the external closed string states. According to conformal symmetry, the points $z_a, z_b$ and $z_c$ are freely fixed to arbitrary points in the complex plane. \par

A mixed string amplitude contains both closed string and open string vertex operators inserted on a disk worldsheet which can be mapped into the upper half plane $H_+=\{z\in \mathbb{C}| \text{Im}(z)\geq 0$\}. Closed string vertex operators are placed onto the bulk while those of open string are inserted along the worldsheet boundary. Unlike the KLT relations, the mixed string amplitudes involving $N_c$ closed strings and $N_o$ open strings can be expressed as linear combinations of $N_o+2N_c$ color-ordered open string amplitudes \cite{Stieberger:2009hq}. At low energies, This relates to Einstein-Yang-Mills amplitudes as a linear combination of pure Yang-Mills amplitudes in the collinear limit 
\cite{Stieberger:2014cea, Stieberger:2015qja}.

In this paper, we will only focus on the mixed amplitudes of ($n-2$) open strings and a single closed string scattering whose expression takes the form \cite{Stieberger:2014cea}
     \begin{align}
          \mathcal{M}_n(1,\dots , n-2;k) = &C_{D_2} (2\pi)^{D}\delta(\sum_{i=1}^{n-2} p_i + k) \int_{\mathcal{I}} \prod_{i=1}^{n-2}dx_i \prod_{1\leq r <s \leq n-2}\abs{x_r-x_s}^{2\alpha'p_r p_s} \nonumber \\
          &\times\int_{H_+} d^2z |z-\bar{z}|^{\frac{1}{2}\alpha'k\cdot k} \prod_{i=1}^{n-2}\abs{x_i-z}^{2\alpha'p_i k}K_n(x_i,z_i,\bar{z}_i) \label{e2}
     \end{align}
     where $ C_{D_2}$ is a normalization constant and the function $K_n$ contains polarization and kinematic factor of the external states. Notice that we used letters $p_i$ and $k$ to denote the momentum of open strings and a closed string respectively. The open string coordinates $x_i$ obey the ordering of the integration region $\mathcal{I}=\{x_i\in\mathbb{R}|x_1<x_2<\ldots<x_{n-2} \}$. The partial amplitudes are associated with a group factor $\Tr{T^{1}T^{2}\dots T^{n-2}}$ with $T^{a}$ being a Chan-Paton factor. This is because the open strings  
     are subject to boundary conditions. The introduction of Chan-Paton factors refers to the fact that each endpoint of the open strings is confined to hyperplanes called D-branes. The factors specify which branes the open strings are attached to. For the sake of computations, we split the closed string momentum $k$ into left- and right-moving spacetime momenta $q_1$ and $q_2$, i.e. $k=q_1+q_2$, which are assumed to be unrelated at first. Therefore, we introduce an integral of the form
      \begin{align}
          \mathcal{F}_n(1,\dots , n-2; q_1,q_2) &=C_{D_2}  \int_{\mathcal{I}} \prod_{i=1}^{n-2}dx_i \prod_{1\leq r <s \leq n-2}\abs{x_r-x_s}^{2\alpha'p_r p_s+n_{rs}}\nonumber \\
         &\times  \int_{H_+} d^2z (z-\bar{z})^{2\alpha'q_1q_2 + n} \prod_{i=1}^{n-2}(x_i-z)^{2\alpha'p_iq_1 + n_i}(x_i - \bar{z})^{2\alpha'p_iq_2 + \bar{n}_i}. \label{e2.1}
     \end{align}
     The integers $n_{rs}$, $n_i$, $\bar{n}_i$, and $n$ are determined by external states. Notice that when $q_1$ = $q_2$ = $k/2$, the integral (\ref{e2.1}) reproduces the mixed string amplitude (\ref{e2}) when the contents of external states are identified.
     
In this paper, we establish a connection between closed and mixed string amplitudes at the tree level. The organization of the paper is as follows: In section two, we establish the relation between closed string and mixed string amplitudes, ranging from 4-point to 6-point ones. Then, a general expression for $n$-point amplitudes is presented in section three. Section 4 is dedicated to a discussion of correction terms pertinent to the results presented in the previous section. Finally, in Section 5, we conclude from our findings, summarizing the key aspects of our work.
\section{Factorization of closed string amplitudes} \label{sec2}
In this section, we would like to formulate expressions for tree-level closed string amplitudes as products of mixed string amplitudes involving open strings and a single closed string. The procedure uses the analytic continuation of complex variables to factorize the closed string amplitudes into the product of those of mixed strings. The calculations in this section will be executed up to the six-point string amplitudes just for the readers to be able to observe patterns of the relations before we generalize them to include arbitrary $n$ strings in the next section.  For simplicity, we begin by discussing the relations that arise at four points.

\subsection{The four-point amplitudes}
    According to the $PSL(2,\mathbb{C})$ symmetry, we choose $z_2=i$, $z_3=-i$ and $z_4=\infty$. The closed string amplitude takes the form
    \begin{align}
        \mathcal{A}_4^{\text{cl}} = 4C_{S^2} \int d^2z \abs{z-i}^{2s_{12}}\abs{z+i}^{2s_{13}}\abs{2i}^{2s_{23}}F_4(z,\bar{z}) \label{e3}
    \end{align}
    where $s_{ij} = \frac{\alpha'}{2}k_i\cdot k_j$. Note that the factor $(2\pi)^D\delta^D(\sum_i k_i)$ has been dropped for convenience. Without loss of generality, we can decompose $F_4$ = $f_4(z)\bar{f}_4(\bar{z})$. By writing $z$ = $x+iy$ and then rotating the contour integral of the variable $y$ from the real axis to the imaginary axis
    \begin{align}
        y \to iy,
    \end{align}
    the integrands transform as follows:
\begin{align}
    |z-i|^{2s_{12}}&\rightarrow (x-y-i)^{s_{12}}(x-y+i)^{s_{12}}, \\
    |z+i|^{2s_{13}}&\rightarrow (x-y-i)^{s_{13}}(x-y+i)^{s_{13}}.
\end{align}
    We then define new variables 
    \begin{equation}
	\xi = x+ y\hspace{1cm}    \text{and}\hspace{1cm} \eta= x-y. \label{e7}
	\end{equation}
Accordingly, the amplitude becomes
 \begin{align}
		\mathcal{A}_4^{\text{cl}} = 4C_{S^2} \int d\xi (\xi + i)^{s_{12}}(\xi-i)^{s_{13}} (-2i)^{s_{23}}\bar{f}_4(\xi) \int d\eta (\eta - i)^{s_{12}}(\eta_1+i)^{s_{13}}(2i)^{s_{23}}f_4(\eta). \label{e8}
	\end{align}
The integrals are exactly those of mixed string amplitudes $\mathcal{F}_4$. If we set $C_{S^2}=(C_{D_2})^2$ and identify the functions $f_4$ and $\bar{f}_4$ as the external-state dependent function $K_4$, we obtain
  \begin{align}
		\mathcal{A}_4^{\text{cl}} = \mathcal{F}_4(1,4;2,3)\widetilde{\mathcal{F}}_4(1,4;3,2) \label{e9}
	\end{align}
providing an expression for four-point closed string amplitudes as a product of mixed string amplitudes in the limit $k_2=k_3$, i.e.
  \begin{align}
		\mathcal{A}_4^{\text{cl}}\Big{\vert}_{k_2=k_3=k} =\mathcal{M}_4(1,4;k)\mathcal{M}_4(1,4;k).  \label{e9-1}
	\end{align}

\subsection{The five-point amplitudes}
    Let's add one more external state into consideration. For a scattering of five strings, when the points $z_3$, $z_4$ and $z_5$ are fixed to the points $i$, $-i$ and $\infty$ respectively in the complex plane, the five-point closed string amplitude reads
    \begin{align}
\mathcal{A}^{\text{c}l}_5 = 4C_{S^2}\int  d^2z_1d^2z_2 & \abs{z_1-i}^{2s_{13}} \abs{z_1+i}^{2s_{14}}\abs{z_2-i}^{2s_{23}} \abs{z_2+i}^{2s_{24}} \abs{2i}^{2s_{34}}\nonumber \\
&\times \abs{z_1-z_2}^{2s_{12}} F_5(z_1,\bar{z}_1,z_2,\bar{z}_2).\label{e10}
\end{align}
We proceed with the same procedure as in the four-point case by rewriting $z_j=x_j+iy_j$ for $j=1,2$ and then the variables $y_j$ are analytically continued to complex variables which allows us to transform
    \begin{align}
        y_j \to ie^{-i\epsilon}y_j \approx iy_j+\epsilon y_j.\label{e11}
    \end{align}
Notice that we rotate the variables $y_j$ from the real axis into almost the imaginary axis to avoid the possible branch points of $\abs{z_1-z_2}^{2s_{12}}$ after the rotations. This changes the integrand
\begin{align}
    \abs{z_j-i}^{2s_{j3}}\rightarrow \quad & (x_j-y_j-i+i\epsilon y_j)^{s_{j3}}(x_j+y_j+i-i\epsilon y_j)^{s_{j3}} \nonumber \\
    \abs{z_j+i}^{2s_{j4}}\rightarrow \quad & (x_j-y_j +i+i\epsilon y_j)^{s_{j4}}(x_j+y_j-i-i\epsilon y_j)^{s_{j4}} \nonumber \\
    \abs{z_1-z_2}^{2s_{12}}\rightarrow \quad & ((x_1-y_1)-(x_2-y_2)+i\epsilon(y_1-y_2))^{s_{12}} \nonumber \\ &\times ((x_1+y_1)-(x_2+y_2)-i\epsilon(y_1-y_2))^{s_{12}} \label{e12}
\end{align}
for $j=1,2$. If we introduce new variables 
\begin{equation}
    \xi_j=x_j+y_j \hspace{1cm}    \text{and}\hspace{1cm} \eta_j= x_j-y_j, \label{new var}
\end{equation}
the integral (\ref{e10}) can be written as
\begin{align}
\mathcal{A}^{\text{cl}}_5 = &4C_{S^2} \int d\xi_1d\eta_1d \xi_2 d\eta_2 (\eta_1-i)^{s_{13}}(\eta_1+i)^{s_{14}}(\eta_2-i)^{s_{23}}(\eta_2+i)^{s_{24}} \nonumber \\ &\times(\xi_1 +i)^{s_{13}}(\xi_1-i)^{s_{14}}(\xi_2 +i)^{s_{23}}(\xi_2-i)^{s_{24}} \abs{2i}^{2s_{34}}\nonumber \\
&\times (\eta_1-\eta_2 +i\epsilon\delta)^{s_{12}} (\xi_1-\xi_2 -i\epsilon\delta)^{s_{12}} f_5(\eta_1,\eta_2)\bar{f}_5(\xi_1,\xi_2) \label{e13}
\end{align}
where $\delta=y_1-y_2$ and $F_5(z_1,\bar{z}_1,z_2,\bar{z}_2)=f_5(z_1,z_2)\bar{f}_5(\bar{z}_1\bar{z}_2)$. The above expression resembles the integral of mixed string amplitudes $\mathcal{F}_5$ albeit a possible phase factor correction due to the intertwining of integrating variables $\xi_j$ and $\eta_j$ of the integrand.

\begin{figure}[t]
           \centering
\subfigure{
           \begin{tikzpicture}
               \draw[<->,thick] (-5,0) -- (-1,0); 
               \draw (-1,1) node {\large{$\eta_1$}};
               \draw (-1.2,1.2) -- (-1.2,0.7);
               \draw (-1.2,0.7) -- (-0.8,0.7);
               \draw[orange] (-3,-0.4) node {$\eta_2$};
                circle (0.05;\fill[black](-3,0) circle (0.05);
                \draw[decoration = {zigzag,segment length = 3mm, amplitude = 1mm}, decorate,orange] (-5,0) -- (-3,0); 
          	\draw(-3,-1) node {$(a)$}; 
                \color{red}
                    \draw[thick](-4.9,0.25)--(-1.1,0.25);
                    \draw[->] (-3,0.25)--(-2.5,0.25);
                    
           \end{tikzpicture} \label{c1a}
}
\subfigure{
\begin{tikzpicture}
     \draw[<->,thick] (1,0) -- (5,0); 
                \draw (4.6,1) node {\large{$\eta_1$}};
               \draw (4.3,1.2) -- (4.3,0.7);
               \draw (4.3,0.7) -- (4.7,0.7);
               \draw [orange](3,0.4) node {$\eta_2$};
               circle (0.05;\fill[black](3,0) circle (0.05);
                \draw[decoration = {zigzag,segment length = 3mm, amplitude = 1mm}, decorate,orange] (1,0) -- (3,0); 
   		\draw(3,-1) node {$(b)$};
      \color{red}
     \draw[thick](1,-0.25)--(5,-0.25);
                    \draw[->](3,-0.25)--(3.5,-0.25);
\end{tikzpicture} \label{c1b}
}

\subfigure{
\begin{tikzpicture}
    \draw[<->,thick] (1,0) -- (5,0); 
                \draw (4.6,1) node {\large{$\xi_1$}};
               \draw (4.3,1.2) -- (4.3,0.7);
               \draw (4.3,0.7) -- (4.7,0.7);
               \draw [orange](3,-0.4) node {$\xi_2$};
               circle (0.05;\fill[black](3,0) circle (0.05);
                \draw[decoration = {zigzag,segment length = 3mm, amplitude = 1mm}, decorate,orange] (1,0) -- (3,0); 
   		\draw(3,-1) node {$(d)$};
                 \color{red}
                    \draw[thick](1,0.25)--(5,0.25);
                    \draw[->](3,0.25)--(3.5,0.25);
\end{tikzpicture} \label{c1c}
}
\subfigure{
           \begin{tikzpicture}
               \draw[<->,thick] (-5,0) -- (-1,0); 
               \draw (-1,1) node {\large{$\xi_1$}};
               \draw (-1.2,1.2) -- (-1.2,0.7);
               \draw (-1.2,0.7) -- (-0.8,0.7);
               \draw[orange] (-3,0.4) node {$\xi_2$};
                circle (0.05;\fill[black](-3,0) circle (0.05);
                \draw[decoration = {zigzag,segment length = 3mm, amplitude = 1mm}, decorate,orange] (-5,0) -- (-3,0); 
          	\draw(-3,-1) node {$(c)$};
                \color{red}
                    \draw[thick](-4.9,-0.25)--(-1.1,-0.25);
                    \draw[->] (-3,-0.25)--(-2.5,-0.25);
                    
           \end{tikzpicture}\label{c1d} 
}
           \caption{Contours of integration for the variable $\eta_1$ in the cases when (a) $\xi_1 > \xi_2$ and (b) $\xi_1< \xi_2$, and contours of integration for the variable $\xi_1$ for the cases where (c) $\eta_1 > \eta_2$ and (d) $\eta_1 < \eta_2$.}
           \label{c1}
       \end{figure}
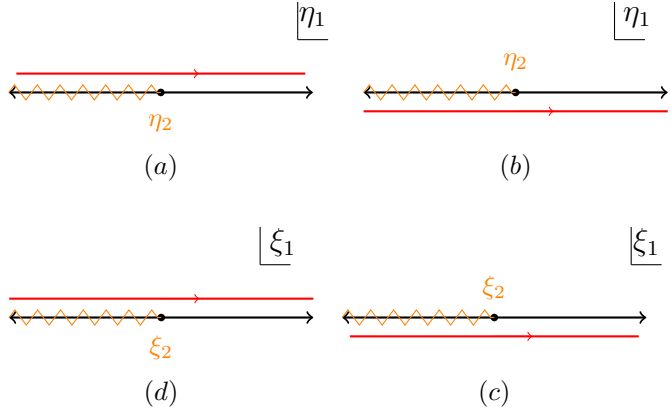

The factor $i\epsilon\delta$ determines deformations of the contours around the branch points, i.e. the points when $\eta_1 \sim \eta_2$ and $\xi_1\sim\xi_2$. One needs to be careful when approaching these points. Let's investigate the behavior of $i\epsilon\delta$ near the branch points. When $\eta_1$ approaches $\eta_2$, $i\epsilon\delta\sim i\frac{\epsilon}{2}(\xi_1-\xi_2)$. This implies that the position of the branch point depends on the sign of $\xi_1-\xi_2$. In case $\xi_1 > \xi_2$, the branch point is located slightly below the real line. On the other hand, if $\xi_1 < \xi_2$, the branch point is slightly above the real line. Consequently, to avoid the branch point, one can slightly shift the $\eta_1$-contour above the real axis when $\xi_1>\xi_2$ and slightly shift the contour down for otherwise. The contours of $\eta_1$ are depicted in figure \ref{c1a} and \ref{c1b}.

A similar pattern also appears for the term $(\xi_1-\xi_2 -i\epsilon\delta)^{s_{12}}$ when $\xi_1$ approaches $\xi_2$. The factor $i\epsilon\delta$ is positive when $\eta_1<\eta_2$ and is negative when $\eta_1>\eta_2$. This leads to a slight shift in the branch point at $\xi_1\sim\xi_2$ to be above and below the real axis of $\xi_1$ for the case $\eta_1>\eta_2$ and $\eta_1<\eta_2$ respectively. As a result, we obtain the $\xi_1$-contours for both cases shown in figure \ref{c1c} and \ref{c1d}.

To proceed, we separate the integral (\ref{e13}) into four terms by inserting the identity
\begin{equation}
    1=(\Theta(\xi_1-\xi_2)+\Theta(\xi_2-\xi_1))(\Theta(\eta_1-\eta_2)+\Theta(\eta_2-\eta_1))
\end{equation}
giving four different integral regions among variables. $\Theta(x-y)$ is a Heaviside step function. It turns out that we can relate the integral in each term to a product of $\mathcal{F}_5$. However, one needs to make sure that the integrand matches with the expression (\ref{e2.1}). For this reason, the following relations are useful to correct the integrand:
\begin{equation}
	(\eta_1-\eta_2)^c =\begin{cases}
					e^{i\pi c}(\eta_2-\eta_1)^c, \hspace{0.7cm} \xi_1 > \xi_2 \\
					e^{-i\pi c}(\eta_2-\eta_1)^c, \hspace{0.5cm} \xi_1 <\xi_2 
                       \end{cases} \label{phase factor eta}
\end{equation} 
when $\text{Re}(\eta_1)<\text{Re}(\eta_2)$ and 
    \begin{equation}
	(\xi_1-\xi_2)^c =\begin{cases}
			
					e^{i\pi c}(\xi_2-\xi_1)^c, \hspace{0.7cm} \eta_1 > \eta_2 .\\
                        e^{-i\pi c}(\xi_2-\xi_1)^c, \hspace{0.5cm} \eta_1 < \eta_2
                       \end{cases}\label{phase factor xi}
\end{equation} 
when $\text{Re}(\xi_1)<\text{Re}(\xi_2)$.

Using the relations (\ref{phase factor eta}) and (\ref{phase factor xi}), one obtains
\begin{align}
\mathcal{A}^{\text{cl}}_5 =  &e^{-i2\pi s_{12}}\mathcal{F}_5(1,2,5;3,4)\widetilde{\mathcal{F}}_5(1,2,5;4,3) +e^{i\pi s_{12}}\mathcal{F}_5(1,2,5;3,4) \widetilde{\mathcal{F}}_5(2,1,5;4,3) \nonumber \\
 &+e^{i\pi s_{12}}\mathcal{F}_5(2,1,5;3,4)\widetilde{\mathcal{F}}_5(1,2,5;4,3)+\mathcal{F}_5(2,1,5;3,4)\widetilde{\mathcal{F}}_5(2,1,5;4,3) \label{e20}
\end{align}
if the constant $C_{S^2}=(C_{D_2})^2$ is fixed. When the $k_3=k_4=k/2$, the integral $\mathcal{F}_5$ reduces to the corresponding mixed string amplitude $\mathcal{M}_5$ with the closed string momentum $k$.

\subsection{The six-point amplitudes}
Before we generalize the relations to an arbitrary number of strings, we would like to consider another non-trivial case regarding six-point string amplitudes, allowing us to observe more patterns in the relations. The expression for an amplitude involving six closed strings reads
\begin{align}
\mathcal{A}^{cl}_6 = 4C_{S^2}\int \prod_{j=1}^{3} d^2z_j  \abs{z_j-i}^{2s_{j4}} \abs{z_j+i}^{2s_{j5}}\abs{2i}^{2s_{45}}F_6(z_j,\bar{z}_j) \prod_{1\leq l<m \leq 3}\abs{z_l-z_m}^{2s_{lm}}
\end{align}
where we choose $z_4=i, z_5=-i$ and $z_6=\infty$ to fix the $PSL(2,\mathbb{C})$ symmetry. We follow the same steps as the previous examples by writing $z_j=x_j+iy_j$ and then rotating $y_j$ using the transformations (\ref{e11}). By introducing the new variables defined 
\begin{equation}
    \xi_j=x_j+y_j \hspace{1cm}    \text{and}\hspace{1cm} \eta_j= x_j-y_j \label{new var2}
\end{equation}
while now $j$ runs from 1 to 3, the amplitude takes the form
\begin{align}
\mathcal{A}^{\text{cl}}_6 = &4C_{S^2} \int \prod_{j=1}^{3}d\eta_j d\xi_j (\eta_j-i)^{s_{j4}}(\eta_j+i)^{s_{j5}}(\xi_j +i)^{s_{j4}}(\xi_j-i)^{s_{j5}} |2i|^{2s_{45}}\nonumber \\
&\times f_6(\xi_j)\bar{f}_6(\eta_j) \prod_{1\leq l<m \leq 3}(\eta_l-\eta_m +i\epsilon\delta_{lm})^{s_{lm}} (\xi_l-\xi_m -i\epsilon\delta_{lm})^{s_{lm}} \label{6 pt amp}
\end{align}
where $\delta_{ij}=y_i-y_j$.

Similar to the four- and five-point cases, the above expression can be rewritten in terms of products of $\mathcal{F}_6$ with appropriate phase factors. The appearance of $i\epsilon\delta_{lm}$ in the integrand dictates how the contours of the integrating variables are deformed. According to the term $(\eta_l-\eta_m+i\epsilon\delta_{lm})^{s_{lm}}$, the $\eta_l$-contour is shifted slightly above the real axis when $\xi_l>\xi_m$ and the contour slightly moves down from the real axis when  $\xi_l<\xi_m$. Likewise, the contour of $\xi_l$ regarding the term  $(\xi_l-\xi_m+i\epsilon\delta_{lm})^{s_{lm}}$ is translated upward from the real axis if $\eta_l>\eta_m$ and is shifted downward from the real axis if $\eta_l<\eta_m$. As the contours are deformed differently depending on how the values of the integrating variables are relative to each other, we will separate the integration domain into 36 pieces based on comparative values of the integration variables by inserting 
\begin{equation}
    1=\sum_{i\neq j\neq k}^3 \Theta(\xi_i-\xi_j)\Theta\xi_j-\xi_k)\sum_{l\neq m\neq n}^3\Theta(\eta_l-\eta_m)\Theta(\eta_m-\eta_n)
\end{equation}
to the expression (\ref{6 pt amp}).

In each term, one can utilize the relations
\begin{equation}
	(\eta_i-\eta_j)^c =\begin{cases}
					e^{i\pi c}(\eta_j-\eta_i)^c, \hspace{0.7cm} \xi_i > \xi_j \\
					e^{-i\pi c}(\eta_j-\eta_i)^c, \hspace{0.5cm} \xi_i <\xi_j 
                       \end{cases} \label{phase factor eta6}
\end{equation} 
when $\text{Re}(\eta_i)<\text{Re}(\eta_j)$, and 
    \begin{equation}
	(\xi_i-\xi_j)^c =\begin{cases}
			
					e^{i\pi c}(\xi_j-\xi_i)^c, \hspace{0.7cm} \eta_i > \eta_j .\\
                        e^{-i\pi c}(\xi_j-\xi_i)^c, \hspace{0.5cm} \eta_i < \eta_j
                       \end{cases}\label{phase factor xi6}
\end{equation} 
when $\text{Re}(\xi_i)<\text{Re}(\xi_j)$ to make sure that the integrand is of the correct form of $\mathcal{F}_6.$ After a careful examination, we obtain the expression
\begin{equation}
    \mathcal{A}_6^{\text{cl}}=\sum_{\sigma,\sigma'} e^{i\pi(\Lambda(\sigma,\sigma')+\Omega(\sigma,\sigma'))}\mathcal{F}_6(\sigma(1,2,3),6;4,5)\widetilde{\mathcal{F}}_6(\sigma'(1,2,3),6;5,4). \label{relation 6pt}
\end{equation}
The expression is summed over the ordering $\sigma$ and $\sigma'$ referring to the order of open string vertex operators in which the objects inside the bracket are to be permuted. The functions $\Lambda(\sigma,\sigma')$ and $\Omega(\sigma,\sigma')$ contain kinematics variables depending on the ordering $\sigma$ and $\sigma'$ respectively. The values of both functions are presented in table \ref{t1}.

\begin{table}[t]
\centering
\begin{minipage}{1.0\textwidth} 
 \begin{tabular}{|c|c|c|c|}
  \hline
  $\sigma$ & $\sigma'$ & $\Lambda(\sigma,\sigma')$ & $\Omega(\sigma,\sigma')$ \\
  \hline
 1,2,3 & 1,2,3 & $-s_{12}-s_{13}-s_{23}$			&$-s_{12}-s_{13}-s_{23}$ \\ 
  \hline 
 1,2,3 & 2,1,3 & $s_{12}-s_{13}-s_{23}$			&$-s_{13}-s_{23}$ \\ 
  \hline 
   1,2,3 & 2,3,1 & $s_{12}+s_{13}-s_{23}$			&$-s_{23}$ \\ 
  \hline 
   1,2,3 & 3,2,1 & $s_{12}+s_{13}+s_{23}$			&$0$ \\ 
  \hline 
   1,2,3 & 3,1,2 & $-s_{12}+s_{13}+s_{23}$			&$-s_{12}$ \\ 
  \hline 
   1,2,3 & 1,3,2 & $-s_{12}-s_{13}+s_{23}$			&$-s_{12}-s_{13}$ \\ 
  \hline 
   2,1,3 & 1,2,3 & $-s_{13}-s_{23}$			&$s_{12}-s_{13}-s_{23}$ \\ 
  \hline 
   2,1,3 & 2,1,3 & $-s_{13}-s_{23}$			&$-s_{13}-s_{23}$ \\ 
  \hline 
   2,1,3 & 2,3,1 & $s_{13}-s_{23}$			&$-s_{23}$ \\ 
  \hline 
   2,1,3 & 3,2,1 & $s_{13}+s_{23}$			&$0$ \\ 
  \hline 
  2,1,3 & 3,1,2 & $s_{13}+s_{23}$			&$s_{12}$ \\ 
  \hline 
   2,1,3 & 1,3,2 & $-s_{13}+s_{23}$			&$s_{12}-s_{13}$ \\ 
  \hline 
   2,3,1 & 1,2,3 & $-s_{23}$			&$s_{12}+s_{13}-s_{23}$ \\ 
  \hline 
   2,3,1 & 2,1,3 & $-s_{23}$			&$s_{13}-s_{23}$ \\ 
  \hline 
   2,3,1 & 2,3,1 & $-s_{23}$			&$-s_{23}$ \\ 
  \hline 
   2,3,1 & 3,2,1 & $s_{23}$			&$0$ \\ 
  \hline 
   2,3,1 & 3,1,2 & $s_{23}$			&$s_{12}$ \\ 
  \hline 
   2,3,1 & 1,3,2 & $s_{23}$			&$s_{12}+s_{13}$ \\ 
  \hline 
   3,2,1 & 1,2,3 & $0$			&$s_{12}+s_{13}+s_{23}$ \\ 
  \hline 
   3,2,1 & 2,1,3 & $0$			&$s_{13}+s_{23}$ \\ 
  \hline 
   3,2,1 & 2,3,1 & $0$			&$s_{23}$ \\ 
  \hline 
  3,2,1 & 3,2,1 & $0$			&$0$ \\ 
  \hline 
   3,2,1 & 3,1,2 & $0$			&$s_{12}$ \\ 
  \hline 
   3,2,1 & 1,3,2 & $0$			&$s_{12}+s_{13}$ \\ 
  \hline 
   3,1,2 & 1,2,3 & $-s_{12}$			&$-s_{12}+s_{13}+s_{23}$ \\ 
  \hline 
   3,1,2 & 2,1,3 & $s_{12}$			&$s_{13}+s_{23}$ \\ 
  \hline 
   3,1,2 & 2,3,1 & $s_{12}$			&$s_{23}$ \\ 
  \hline 
   3,1,2 & 3,2,1 & $s_{12}$			&$0$ \\ 
  \hline 
   3,1,2 & 3,1,2 & $-s_{12}$			&$-s_{12}$ \\ 
  \hline 
   3,1,2 & 1,3,2 & $-s_{12}$			&$-s_{12}+s_{13}$ \\ 
  \hline 
   1,3,2 & 1,2,3 & $-s_{12}-s_{13}$			&$-s_{12}-s_{13}+s_{23}$ \\ 
  \hline 
   1,3,2 & 2,1,3 & $s_{12}-s_{13}$			&$-s_{13}+s_{23}$ \\ 
  \hline 
   1,3,2 & 2,3,1 & $s_{12}+s_{13}$			&$s_{23}$ \\ 
  \hline 
   1,3,2 & 3,2,1 & $s_{12}+s_{13}$			&$0$ \\ 
  \hline 
   1,3,2 & 3,1,2 & $-s_{12}+s_{13}$			&$-s_{12}$ \\ 
  \hline 
   1,3,2 & 1,3,2 & $-s_{12}-s_{13}$			&$-s_{12}-s_{13}$ \\ 
  \hline 
 \end{tabular}
 \end{minipage}
 \vspace{10pt} 
\caption{Values of functions $\Lambda(\sigma,\sigma')$ and $\Omega(\sigma,\sigma')$ corresponding to the ordering $\sigma$ and $\sigma'$.}
 \label{t1}
\end{table}

\section{Relations between closed and mixed string amplitudes}\label{sec3}
In this section, we would like to generalize the results formulated in the previous section to include an arbitrary number of strings. To do that, we repeat the same procedure used in the last section to rewrite the expressions for $n$-point closed string amplitudes,
\begin{align}
    \mathcal{A}^{\text{cl}}_n = &4C_{S^2}\int \prod_{j=1}^{n-3} d^2z_j  \abs{z_j-i}^{2s_{j,n-2}} \abs{z_j+i}^{2s_{j,n-1}}\abs{2i}^{2s_{n-2,n-1}}F_n(z_j,\bar{z}_j) \nonumber \\
    &\prod_{1\leq l<m \leq n-3}\abs{z_l-z_m}^{2s_{lm}}
\end{align}
with $z_{n-2}=i, z_{n-1}=-i$ and $z_n=\infty$, into products of mixed string amplitudes. Again, we rewrite $z_j=x_j+iy_j$ and then transform the variable $y_j$ to $ie^{-i\epsilon}y_j$. It yields
\begin{align}
\mathcal{A}^{\text{cl}}_n = &4C_{S^2}\int \prod_{j=1}^{n-3}d\eta_j d\xi_j (\eta_j-i)^{s_{j,n-2}}(\eta_j+i)^{s_{j,n-1}}(\xi_j +i)^{s_{j,n-2}}(\xi_j-i)^{s_{j,n-1}} \nonumber \\
&f_n(\xi_j)\bar{f}_n(\eta_j) \prod_{1\leq l<m \leq n-3}(\eta_l-\eta_m +i\epsilon\delta_{lm})^{s_{lm}} (\xi_l-\xi_m -i\epsilon\delta_{lm})^{s_{lm}}|2i|^{2s_{n-2,n-1}} \label{npt amp}
\end{align}
where the variables $\xi_j$ and $\eta_j$ are defined in (\ref{new var2}).

By a careful examination of branch cuts and contour deformations, we can write the general expression for the closed string amplitudes as 
\begin{align}
    \mathcal{A}_n^{\text{cl}} =  &\sum_{\sigma,\sigma'} \mathcal{P}(\sigma(1,2,3,\dots,n-3)|\sigma'(1,2,3,\dots,n-3)) \nonumber \\
   & \times\mathcal{F}_n(\sigma(1,2,3,\dots,n-3),n ; n-2,n-1)
   \nonumber \\
   &\times \widetilde{\mathcal{F}}_n(\sigma'(1,2,3,\dots,n-3),n ;n-1,n-2) \label{e30}
\end{align}
with the help of the relations (\ref{phase factor eta6}) and (\ref{phase factor xi6}). This resembles the relation (\ref{relation 6pt}) where the function $\mathcal{P}(\sigma|\sigma')$ encapsulates all the phase factors appearing when correcting the integrand of (\ref{npt amp}). We define the function $\mathcal{P}(\sigma|\sigma')$ by introducing two new functions $\theta(i_j,i_k)$ and $\beta(i_j,i_k)$. The function $\theta(i_j,i_k)$ is defined as
\begin{equation}
	\theta(i_p,i_q) =\begin{cases}
					1 \hspace{0.5 cm} ; \hspace{0.5 cm} (i_p,i_q)\hspace{0.1 cm} \text{has the same ordering as set}\hspace{0.1 cm} I_n \\
					0 \hspace{0.5 cm}; \hspace{0.5 cm} (i_p,i_q)\hspace{0.1 cm} \text{has the opposite ordering as set} \hspace{0.1 cm}I_n\\
                       \end{cases} \label{e27}
\end{equation} 
where the set $I_n$ are given by
\begin{equation}
    I_n = \{1,2,3,\dots,n-3\}.
\end{equation}
For example, $\theta(1,2)$ = 1, $\theta(2,1)$ = 0, $\theta(4,7)$ = 0, etc. The function $\beta(i_j,i_k)$ is defined as
\begin{equation}
	\beta (i_p,i_q | j_r,j_s) =\begin{cases}
					-1 \hspace{0.5 cm} ; \hspace{0.5 cm} (i_p,i_q) \hspace{0.1 cm}\text{and}\hspace{0.1 cm} (j_r,j_s)\hspace{0.1 cm} \text{have the same ordering}, \\
                     +1 \hspace{0.5 cm} ; \hspace{0.5 cm} (i_p,i_q) \hspace{0.1 cm}\text{and}\hspace{0.1 cm} (j_r,j_s)\hspace{0.1 cm} \text{have the opposite ordering },\\
					0 \hspace{0.8 cm}; \hspace{0.5 cm} (i_p,i_q) \hspace{0.1 cm}\text{and}\hspace{0.1 cm} (j_r,j_s)\hspace{0.1 cm} \text{are different. }\\
                       \end{cases} \label{e28}
\end{equation} 
For example, $\beta(1,2|1,2)$ = -1, $\beta(1,2|2,1)$ = +1, $\beta(1,2|1,3)$ = 0, etc. We then construct the function $\mathcal{P}(\sigma|\sigma')$ to be
\begin{align}
    \mathcal{P}(i_1,i_2,\dots,i_k|j_1,j_2,\dots,j_k) =&\exp \Big{\{}i\pi\big{(}\sum_{p<q}^k\sum_{r<s}^k\theta(i_p,i_q)\beta(i_p,i_q|j_r,j_s)s_{i_pi_q} \nonumber \\   &+\theta(j_r,j_s)\beta(j_r,j_s|i_p,i_q)s_{j_rj_s}\big{)}\Big{\}}.\label{e29}
\end{align}

The expression (\ref{e30}) contains $(n-3)!\times(n-3)!$ terms in the summation regarding the permutations of $n-3$ objects inside the ordering $\sigma$ and $\sigma'$. In the case where $k_{n-2}=k_{n-1}=k/2$, the relation reduces to
\begin{align}
    \mathcal{A}_n^{\text{cl}}\Big|_{k_{n-2}=k_{n-1}=k/2} =  &\sum_{\sigma,\sigma'} \mathcal{P}(\sigma(1,2,3,\dots,n-3)|\sigma'(1,2,3,\dots,n-3)) \nonumber \\
   & \times\mathcal{M}_n(\sigma(1,2,3,\dots,n-3),n ; k)
   \nonumber \\
   &\times \widetilde{\mathcal{M}}_n(\sigma'(1,2,3,\dots,n-3),n ;k). \label{e30-1}
\end{align}
Remind that the polarizations of closed strings $\chi_i$ corresponding to the physical states of $\mathcal{A}_n^{\text{cl}}$ are decomposed into open string polarizations $\zeta_i$ and $\tilde{\zeta}_i$ regarding the amplitudes $\mathcal{M}_n$ and $\widetilde{\mathcal{M}}_n$ respectively through $\chi_i=\zeta_i\otimes \tilde\zeta_i$. However, the polarizations of the closed string in the mixed amplitudes $\mathcal{M}_n$ and $\widetilde{\mathcal{M}}_n$ read $\zeta_{n-2}\otimes \zeta_{n-1}$ and $\tilde\zeta_{n-2}\otimes \tilde\zeta_{n-1}$ respectively.
More detail can be found in appendix \ref{A1}.

\section{Correction terms}\label{sec12}

Although we have formulated the relations between closed string amplitudes and those of mixed string amplitudes presented in (\ref{e30}) and $(\ref{e30-1})$, the calculations require Wick rotations of the variables $y_j$ via (\ref{e11}). This step naively assumes no poles or branch points when implementing the contour integral. In this section, we will take a good care of such possible points that can exist in the bulk of the complex plane resulting in a possible correction to the proposed relations. For simplicity, we will content ourselves to consider only the four-point relation yet provide comments on the higher-point cases at the end of the section.

 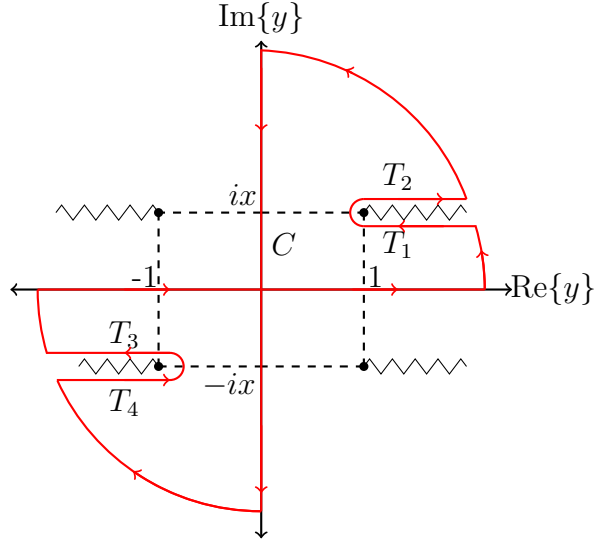
\begin{figure}[t]
         \centering
         \begin{tikzpicture}[scale=0.6]
        \draw[<->,thick] (-5.5,0) -- (5.5,0); 
         \draw[<->,thick] (0,-5.5) -- (0,5.5); 
          \draw (0.5,1) node {\large{$C$}};
            \draw (3,1) node {\large{$T_1$}};
            \draw (3,2.5) node {\large{$T_2$}};
            \draw (-3,-1) node {\large{$T_3$}};
            \draw (-3,-2.5) node {\large{$T_4$}};
            \draw (-0.4,2.1) node {\large{$ix$}};
            \draw (-0.7,-2.1) node {\large{$-ix$}};
            \draw (2.5,0.28) node {\large{1}};
            \draw (-2.55,0.28) node {\large{-1}};
            \draw (0,6) node {\large{$\Im{y}$}};
            \draw (6.4,0) node {\large{$\Re{y}$}};
            circle (0.1);\fill[black](2.25,1.7) circle (0.1);
            circle (0.1);\fill[black](-2.25,1.7) circle (0.1);
            circle (0.1);\fill[black](-2.25,-1.7) circle (0.1);
            circle (0.1);\fill[black](2.25,-1.7) circle (0.1);
            \draw[thick,dashed] (2.25,1.7)--(2.25,0);
            \draw[thick,dashed] (-2.25,1.7)--(-2.25,0);
            \draw[thick,dashed] (0,1.7)--(2.25,1.7);
            \draw[thick,dashed] (0,1.7)--(-2.25,1.7);
            \draw[thick,dashed] (-2.25,-1.7)--(-2.25,0);
            \draw[thick,dashed] (2.25,-1.7)--(2.25,0);
            \draw[thick,dashed] (-2.25,-1.7)--(0,-1.7);
            \draw[thick,dashed] (2.25,-1.7)--(0,-1.7);
            \draw[decoration = {zigzag,segment length = 3mm, amplitude = 1mm}, decorate] (2.25,1.7) -- (4.5,1.7);
            \draw[decoration = {zigzag,segment length = 3mm, amplitude = 1mm}, decorate] (-4.5,1.7) -- (-2.25,1.7);
            \draw[decoration = {zigzag,segment length = 3mm, amplitude = 1mm}, decorate] (-4,-1.7) -- (-2.25,-1.7);
            \draw[decoration = {zigzag,segment length = 3mm, amplitude = 1mm}, decorate] (2.25,-1.7) -- (4.5,-1.7);
         \color{red}
         \draw[thick] (4.9,0) arc (0:16.5:5);
         \draw[->,thick] (4.9,0) arc (0:10:5); 
         \draw[thick] (4.5,2) arc (20:88:5); 
         \draw[->,thick] (2,4.78) arc (60:62:5); 
         \draw[thick] (2.25,2)-- (4.5,2); 
         \draw[->,thick] (3,2) -- (4,2);
         \draw[thick] (2.25,1.4)-- (4.7,1.4); 
         \draw[->,thick] (4,1.4) -- (3,1.4);
         \draw[thick] (2.25,2) arc (90:270:0.3);
         \draw[thick] (0,5.3)-- (0,0);
         \draw [->,thick] (0,4.9)-- (0,3.5);
         
         \draw [thick](-4.9,0) arc (180:196.5:5);
         \draw [->,thick](0,-4.9) arc (270:235:4.9); 
         \draw[thick] (0,-4.9) arc (270:204.3:4.9); 
         \draw [thick](-2,-1.4)-- (-4.7,-1.4);
         \draw [->,thick](-2,-1.4)-- (-3,-1.4);
         \draw[thick] (-2,-2)-- (-4.48,-2);
         \draw[->,thick]  (-3,-2)--(-2,-2);
         \draw[thick] (-2,-2) arc (270:450:0.3);
         \draw [thick] (0,0) -- (0,-4.9);
         \draw [->,thick] (0,-3) -- (0,-4.5); 
         \draw[thick] (-4.9,0) -- (4.9,0);
         \draw[->,thick] (-3,0) -- (-2,0);
         \draw[->,thick] (2,0) -- (3,0);
    \end{tikzpicture}
    \vspace{10pt}
         \caption{Contour integral in the complex $y$ plane.}
         \label{c9}
     \end{figure} 

It is clear that the expression of the four-point closed string amplitude (\ref{e3}) contains the terms $|z-i|^{2s_{12}}$ and $|z+i|^{2s_{13}}$ which implies an existence of branch points at $y=\pm(1 \pm ix)$ inside the worldsheet. To transform $y$ into an imaginary axis, one needs to make sure that the $y$-contour avoids intersecting any cuts in the complex plane. To address this, we define the contour of $y$ as shown in figure \ref{c9}. Following this contour, additional terms emerge from the integration around the infinite tubes, namely paths $T_1, T_2, T_3$, and $T_4$. As there are no poles within the contour, the variable $y$ undergoes transformation using the Cauchy theorem:
\begin{align}
   \Big( \int_{-\infty}^\infty + \int_{T_1+T_2}+\int_{T_3+T_4}+\int_{C} \Big)dy \ I(y)=0.
\end{align}
Here, $I(y)$ represents an arbitrary analytic function. It is important to note that we disregard contributions from the integral along infinite arcs, assuming that the function $I(y)\rightarrow 0$ when $|y|\rightarrow \infty$ which holds true for our integrand. Therefore, the tree-level four-point closed string amplitude becomes
\begin{align}
    \mathcal{A}_4^{\text{cl}} = \mathcal{F}_4(1,4;2,3)\widetilde{\mathcal{F}}_4(1,4;3,2) +8iC_{S^2}\int_{-\infty}^\infty dx \Big(\int_{T_1+T_2}+\int_{T_3+T_4}\Big)dy \ \mathcal{I}(x,y)\label{additional term}
\end{align}
where 
\begin{align}
    \mathcal{I}(x,y) = (x+iy-i)^{s_{12}}(x-iy+i)^{s_{12}}(x+iy+i)^{s_{13}}(x-iy-i)^{s_{13}}|2i|^{2s_{23}}F_4(x,y).
\end{align}
The first term of (\ref{additional term}) emerges from the contour integration along the imaginary line (path $C$) we already discussed. 

Now let's delve into the additional terms resulting from the integration along the infinite tubes by investigating the following integrals:
\begin{align}
    \int_{T_1+T_2} dy \ \mathcal{I}(x,y) \qquad \text{and} \qquad \int_{T_3+T_4} dy \ \mathcal{I}(x,y) 
\end{align}
where we will denote these integrals $A_1$ and $A_2$ respectively. For convenience, we set $F_4=1$. To compute, we will examine the additional integral in two cases: when $x>0$ and $x<0$.

   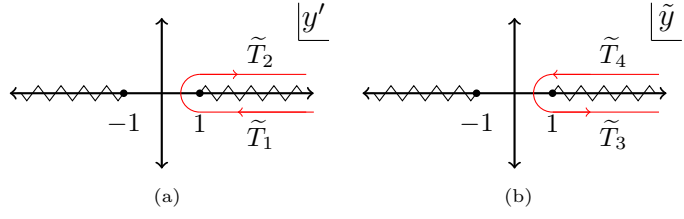
\begin{figure}[t]
           \centering
        \subfigure[]{
           \begin{tikzpicture}
               \draw[<->,thick] (-5,0) -- (-1,0); 
               \draw[<->,thick] (-3,-1) -- (-3,1); 
               \draw (-1,1) node {\large{$y'$}};
               \draw (-1.2,1.2) -- (-1.2,0.7);
               \draw (-1.2,0.7) -- (-0.8,0.7);
               \draw (-2.5,-0.4) node {$1$};
               \draw (-3.5,-0.4) node {$-1$};
                circle (0.05;\fill[black](-2.5,0) circle (0.05);
                circle (0.05;\fill[black](-3.5,0) circle (0.05);
                \draw[decoration = {zigzag,segment length = 3mm, amplitude = 1mm}, decorate] (-2.5,0) -- (-1.1,0);
                \draw[decoration = {zigzag,segment length = 3mm, amplitude = 1mm}, decorate] (-3.5,0) -- (-4.9,0); 
                \draw (-1.7,-0.55) node {$\widetilde{T}_1$};
                \draw (-1.7,0.55) node {$\widetilde{T}_2$};
                \color{red}
                    \draw(-2.5,0.25)--(-1.1,0.25);
                    \draw[->](-2.5,0.25)--(-2,0.25);
                    \draw(-2.5,-0.25)--(-1.1,-0.25);
                    \draw[->](-1,-0.25)--(-2,-0.25);
                    \draw (-2.5,0.25) arc (90:270:0.25); 
           \end{tikzpicture}\label{c10a}
           }
           \subfigure[]{
           \begin{tikzpicture}
               \draw[<->,thick] (-5,0) -- (-1,0); 
               \draw[<->,thick] (-3,-1) -- (-3,1); 
               \draw (-1,1) node {\large{$\tilde{y}$}};
               \draw (-1.2,1.2) -- (-1.2,0.7);
               \draw (-1.2,0.7) -- (-0.8,0.7);
               \draw (-2.5,-0.4) node {$1$};
               \draw (-3.5,-0.4) node {$-1$};
                circle (0.05;\fill[black](-2.5,0) circle (0.05);
                circle (0.05;\fill[black](-3.5,0) circle (0.05);
                \draw[decoration = {zigzag,segment length = 3mm, amplitude = 1mm}, decorate] (-2.5,0) -- (-1.1,0);
                \draw[decoration = {zigzag,segment length = 3mm, amplitude = 1mm}, decorate] (-3.5,0) -- (-4.9,0); 
                
                \draw (-1.7,-0.55) node {$\widetilde{T}_3$};
                \draw (-1.7,0.55) node {$\widetilde{T}_4$};
                \color{red}
                    \draw(-2.5,0.25)--(-1.1,0.25);
                    \draw[->](-2,0.25)--(-2.5,0.25);
                    \draw(-2.5,-0.25)--(-1.1,-0.25);
                    \draw[->](-2.5,-0.25)--(-2,-0.25);
                    \draw (-2.5,0.25) arc (90:270:0.25); 
           \end{tikzpicture} \label{c10b}
           }
           \caption{Contour along the infinite tubes of (a) the variable $y'$ and (b) the variable $\Tilde{y}$.}
           \label{c10}
       \end{figure} 

For the first case $x>0$, by changing the variable $y$ to $y'$ where
\begin{equation}
    y=y'+ix,
\end{equation}
the integral $A_1$ becomes
\begin{align}
    A_1\Big|_{x>0} = &i^{(s_{12}+s_{13})}4^{s_{23}}\Big(\int_{\widetilde{T}_1}+\int_{\widetilde{T}_2}\Big)dy'(y'-1)^{s_{12}}(x-iy'+x+i)^{s_{12}}\nonumber \\ &\times(y'+1)^{s_{13}}(x-iy'+x-i)^{s_{13}}.
\end{align}
The contour paths $\widetilde{T}_1$ and $\widetilde{T}_2$ of the contour $y'$ are illustrated in figure \ref{c10a}. Using the relation
       \begin{equation}
				(y'-1)^c=\begin{cases}
					\abs{y'-1}^c e^{2\pi i c} ;\hspace{0.3cm} \text{for path }\widetilde{T}_1 \\
                        \abs{y'-1}^c ; \hspace{1cm} \text{for path }\widetilde{T}_2, 
				\end{cases} \label{e36}
		\end{equation} 
one can write 
  \begin{align}
      A_1\Big|_{x>0} = &-(2i) i^{(s_{12}+s_{13})}4^{s_{23}} \sin{(\pi s_{12})}e^{i\pi s_{12}}\int^{\infty}_{1}dy'\abs{y'-1}^{s_{12}}\abs{y'+1}^{s_{13}}\nonumber \\ &\times(x-iy'+x+i)^{s_{12}}(x-iy'+x-i)^{s_{13}}. \label{e37}
  \end{align}
Likewise, one can compute $A_2$ for the case $x>0$ to be
\begin{align}
    A_2\Big|_{x>0} = &i^{(s_{12}+s_{13})}4^{s_{23}}\Big(\int_{\widetilde{T}_3}+\int_{\widetilde{T}_4}\Big)d\tilde{y}(\tilde{y}-1)^{s_{13}}(x-i\tilde{y}+x+i)^{s_{13}}\nonumber \\ &\times(\tilde{y}+1)^{s_{12}}(x-i\tilde{y}+x-i)^{s_{12}}
\end{align}
where we define
\begin{equation}
    y=-\tilde{y}-ix.
\end{equation}
The figure \ref{c10b} illustrates paths $\widetilde{T}_3$ and $\widetilde{T}_4$ of the $\tilde{y}$-contour along the infinite tube. Again, we then apply the relation
       \begin{equation}
				(\tilde{y}-1)^c=\begin{cases}
					\abs{\tilde{y}-1}^c e^{2\pi i c} ;\hspace{0.3cm} \text{for path }\widetilde{T}_3 \\
                        \abs{\tilde{y}-1}^c ; \hspace{1cm} \text{for path }\widetilde{T}_4 
				\end{cases} \label{e41}
		\end{equation} 
to rewrite $A_2$ in the form
    \begin{align}
      A_2\Big|_{x>0} = &-(2i) i^{(s_{12}+s_{13})}4^{s_{23}}\sin{(\pi s_{13})}e^{i\pi s_{13}}\int^{\infty}_{1}d\tilde{y}\abs{\tilde{y}-1}^{s_{13}}\abs{\tilde{y}+1}^{s_{12}}\nonumber \\ &\times(x-i\tilde{y}+x+i)^{s_{13}}(x-i\tilde{y}+x-i)^{s_{12}}. \label{e38}
  \end{align}

Without much effort, one can compute expressions for $A_1$ and $A_2$ for the latter case where $x<0$ and find out that
\begin{equation}
    \int_0^\infty dx\Big( A_1\Big|_{x>0}+A_2\Big|_{x>0}\Big)=\int_{-\infty}^0 dx\Big( A_1\Big|_{x<0}+A_2\Big|_{x<0}\Big).
\end{equation}
At last, the four-point closed string amplitude (\ref{additional term}) takes the form
\begin{align}
    \mathcal{A}_4^{\text{cl}} = \mathcal{F}_4(1,4;2,3)\widetilde{\mathcal{F}}_4(1,4;3,2) +32C_{S^2}i^{(s_{12}+s_{13})}(2^{s_{23}})\Big( \mathcal{B}(s_{12},s_{13})+\mathcal{B}(s_{13},s_{12})\Big)\label{additional term2}
\end{align}
where
\begin{align}
    \mathcal{B}(s_{12},s_{13})= &\sin{(\pi s_{12})}e^{i\pi s_{12}}\int_0^\infty dx\int^{\infty}_{1}dy'\abs{y-1}^{s_{12}}\abs{y+1}^{s_{13}}\nonumber \\ &\times(x-iy+x+i)^{s_{12}}(x-iy+x-i)^{s_{13}}2^{s_{23}}. \label{additional fn}
\end{align}
Remember that the order of the arguments of $\mathcal{B}(x,y)$ matters, i.e. $\mathcal{B}(x,y)\neq \mathcal{B}(y,x)$.

\begin{figure}[t]
    \centering
\subfigure[]{
\begin{tikzpicture}
         \draw[<->,thick] (-3,0) -- (2,0); 
         \draw (2.6,0) node {$\Re{x}$}; 
         \draw (-2,3.7) node {$\Im{x}$};
         \draw[<->,thick] (-2,-0.5) -- (-2,3.5); 
         \draw (-2.2,1) node {$i$};
          circle (0.05;\fill[black](-2,1) circle (0.05);
         \draw (-2.2,2) node {$iy$};
         circle (0.05;\fill[black](-2,2) circle (0.05);
         \draw[blue,thick] (-1.9,0) -- (1.8,0); 
          \draw[blue,thick] (1.8,0) arc (0:99:3.2);
           \draw[blue,thick] (-1.9,3.15) -- (-1.9,2.2); 
           \draw[blue,thick] (-1.9,1.8) arc (270:450:0.2);
            \draw[blue,thick] (-1.9,1.8) -- (-1.9,1.2); 
          \draw[blue,thick] (-1.9,0.8) arc (270:450:0.2);
            \draw[blue,thick] (-1.9,0.8) -- (-1.9,0); 
    \end{tikzpicture}   \label{c4a}
}
\subfigure[]{
\begin{tikzpicture}
         \draw[<->,thick] (2,1) -- (8.2,1); 
         \draw (8,2) node {$x$};
         \draw[<->,thick] (5,0) -- (5,2); 
         \draw (6,0.8) node {$1$};
         \draw (5,-1) node {};
          circle (0.05;\fill[black](6,1) circle (0.05);
          \draw[decoration = {zigzag,segment length = 3mm, amplitude = 1mm}, decorate,green] (6,1) -- (8,1);
         \draw (7,0.8) node {$y$};
         circle (0.05;\fill[black](7,1) circle (0.05);
         \draw[decoration = {zigzag,segment length = 3mm, amplitude = 1mm}, decorate,orange] (7,1) -- (8,1);
         \draw (4,0.8) node {$-1$};
         circle (0.05;\fill[black](4,1) circle (0.05);
         \draw[decoration = {zigzag,segment length = 3mm, amplitude = 1mm}, decorate,pink] (2,1) -- (4,1);
         
         \draw[blue,thick] (5,0.7) -- (5.8,0.7);
         \draw[blue,thick] (5.8,0.7) arc (180:360:0.2);
         \draw[blue,thick] (6.2,0.7) -- (6.8,0.7);
         \draw[blue,thick] (6.8,0.7) arc (180:360:0.2);
         \draw[blue,thick] (7.2,0.7) -- (7.9,0.7); 
         \draw (6.5,0.5) node {$L$}; 
         
         \draw[red,thick] (5,1.3) -- (5.8,1.3);
         \draw[red,thick] (6.2,1.3) arc (0:180:0.2);
         \draw[red,thick] (6.2,1.3) -- (6.8,1.3);
         \draw[red,thick] (7.2,1.3) arc (0:180:0.2);
         \draw[red,thick] (7.2,1.3) -- (7.9,1.3);
         \draw (6.5,1.5) node {$L^*$}; 
        \end{tikzpicture} \label{c4b}
}
    \caption{(a) Contour of integration providing a Wick rotation of variable $x$ and (b) contour path of the variable $x$ after the rotation}
    \label{f14}
\end{figure}
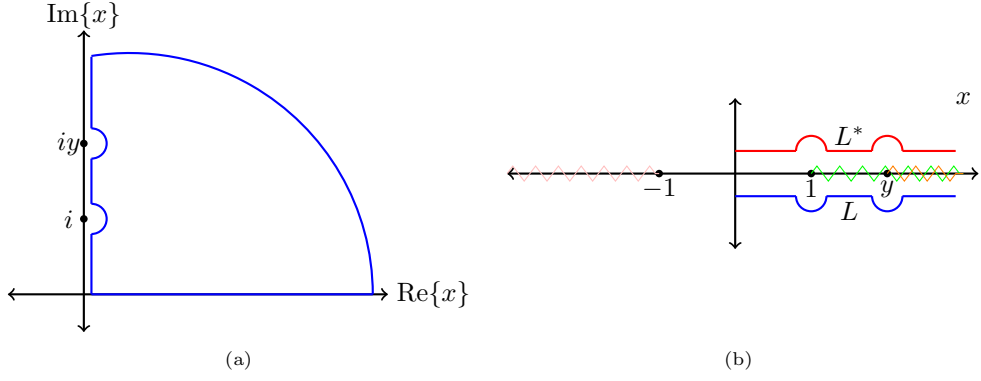

Furthermore, we noticed that the integral $\mathcal{B}(x,y)$ can be related to color-ordered open string amplitudes. This can be done by implementing binomial expansions to the expression (\ref{additional fn}), giving
\begin{align}
    \mathcal{B}(s_{12},s_{13}) =&\sin{(\pi s_{12})}e^{i\pi s_{12}}\sum_{a,b=0}^{\infty} \binom{s_{12}}{a}\binom{s_{13}}{b} \int_{1}^{\infty}dy \ \abs{y-1}^{s_{12}}\abs{y+1}^{s_{13}} \nonumber \\
    &\times\int_{0}^{\infty}dx \ (x+i)^{s_{12}-a}(x-iy)^{a+b}(x-i)^{s_{13}-b}2^{s_{23}}. \label{e56}
\end{align}
If we sent $x$ $\to$ $ie^{-i\epsilon}x$, $\mathcal{B}(s_{12},s_{13})$ becomes
\begin{align}
    \mathcal{B}(s_{12},s_{13}) = &i\sin{(\pi s_{12})}e^{i\frac{\pi}{2}(3s_{12}+s_{13})}\sum_{a,b=0}^{\infty} \binom{s_{12}}{a}\binom{s_{13}}{b}  \int_{1}^{\infty}dy \abs{y-1}^{s_{12}}\abs{y+1}^{s_{13}} \nonumber \\
    &\times\int_{L}dx(x+1)^{s_{12}-a}(x-y)^{a+b}(x-1)^{s_{13}-b} 2^{s_{23}}. \label{e60}
\end{align}
The parameter $\epsilon$ was there to avoid intersecting branch points on the imaginary axis. Figure \ref{c4a} shows the contour of integration we followed so that the variable $x$ is rotated to the imaginary axis. After the rotation, the new variable $x$ (the rotated one) follows the contour $L$ shown in figure \ref{c4b}. Now, we would like to refer $\mathcal{B}(s_{12},s_{13})$ as $\mathcal{B}_{L}(s_{12},s_{13})$ to signify the path of integration. By assigning the branch cuts to be along the real axis shown in figure \ref{c4b}, one can write
\begin{align}
   \mathcal{B}_L(s_{12},s_{13}) = &i\sin{(\pi s_{12})}e^{i\frac{\pi}{2}(3s_{12}+s_{13})}\sum_{a,b=0}^{\infty} \binom{s_{12}}{a}\binom{s_{13}}{b}  \int_{1}^{\infty}dy \abs{y-1}^{s_{12}}\abs{y+1}^{s_{13}} \nonumber \\
    &\times e^{\pi i(s_{13}+a)}\bigg(\int_{0}^{1}dx+e^{\pi i (s_{13}-b)}\int_{1}^{y}dx+e^{\pi i(s_{13}+a)}\int_{y}^{\infty}dx\bigg)\nonumber \\
    &\times\abs{x+1}^{s_{12}-a}\abs{x-y}^{a+b}\abs{x-1}^{s_{13}-b}2^{s_{23}}. \label{e61}
\end{align}
It is not hard to see that the integrals that involve the integration region $x\in(1,y)$ and $x\in(y,\infty)$ (the last two terms in the second line of (\ref{e61})) are the integral forms of color-ordered open string amplitudes considering that the open string vertex positions are chosen to be at $-1, 1$ and $\infty$ while gauge fix. According to \cite{Green:2012oqa}, the integral describing tree-level five-point partial string amplitude is
\begin{align}
    \mathcal{I}^{\text{op}}_5(2,3,1,4,5;\{n_{ij}\})=&\int_1^\infty dx_4 \int_{1}^{x_4} dx_1 (x_1+1)^{s_{12}+n_{12}}(x_1-1)^{s_{13}+n_{13}} \nonumber \\
    &\times(x_4+1)^{s_{24}+n_{24}}(x_4-1)^{s_{34}+n_{34}}(x_4-x_1)^{s_{14}+n_{14}}2^{s_{23}} \label{open integral}
\end{align}
with a set of integers $n_{ij}$. The three points of open string vertices are fixed to be $x_2=-1$, $x_3=1$, and $x_5=\infty$. Note that if one would like to identify the integral (\ref{open integral}) to the open string amplitude, the integers $n_{ij}$ are to be fixed to certain values associated with external string states.

To get rid of the terms involving the integration of $x$ from 0 to 1 of (\ref{e61}), we consider the function $\mathcal{B}_{L^*}(s_{12},s_{13})$ whose path of integration $L^*$ is defined as shown in figure \ref{c4b}. The function $\mathcal{B}_{L^*}(s_{12},s_{13})$ is in the form
\begin{align}
   \mathcal{B}_{L^*}(s_{12},s_{13}) = &i\sin{(\pi s_{12})}e^{i\frac{\pi}{2}(3s_{12}+s_{13})}\sum_{a,b=0}^{\infty} \binom{s_{12}}{a}\binom{s_{13}}{b}  \int_{1}^{\infty}dy \abs{y-1}^{s_{12}}\abs{y+1}^{s_{13}} \nonumber \\
    &\times e^{\pi i(s_{13}+a)}\bigg(\int_{0}^{1}dx+e^{-\pi i (s_{13}-b)}\int_{1}^{y}dx+e^{-\pi i(s_{13}+a)}\int_{y}^{\infty}dx\bigg)\nonumber \\
    &\times\abs{x+1}^{s_{12}-a}\abs{x-y}^{a+b}\abs{x-1}^{s_{13}-b}2^{s_{23}}. \label{e61-1}
\end{align}
If we subtract $\mathcal{B}_{L}(s_{12},s_{13})$ by $\mathcal{B}_{L^*}(s_{12},s_{13})$, one yields
\begin{align}
    &i\sin{(\pi s_{12})}e^{i\frac{\pi}{2}(3s_{12}+s_{13})}\sum_{a,b=0}^{\infty} \binom{s_{12}}{a}\binom{s_{13}}{b}  \int_{1}^{\infty}dy \abs{y-1}^{s_{12}}\abs{y+1}^{s_{13}} \nonumber \\
    &\times 2i e^{\pi i(s_{13}+a)} \bigg(\sin{(\pi(s_{13}-b))}\int_{1}^{y}dx+\sin{(\pi(s_{13}+a))}\int_{y}^{\infty}dx\bigg)
    \nonumber \\
    &\times\abs{x+1}^{s_{12}-a}\abs{x-y}^{a+b}\abs{x-1}^{s_{13}-b} \label{e63}
\end{align}
which can be written in terms of $\mathcal{I}^{\text{op}}_5$ as
\begin{align}
    \mathcal{B}_{L}(s_{12},s_{13})&-\mathcal{B}_{L^*}(s_{12},s_{13})=-2\sin{(\pi s_{12})}e^{i\frac{3}{2}\pi(s_{12}+s_{13})}\sum_{a,b=0}^{\infty} \binom{s_{12}}{a}\binom{s_{13}}{b} (-1)^a \nonumber \\
    &\times\bigg( \sin{(\pi(s_{13}-b))} \mathcal{I}^{\text{op}}_5(2,3,1,4,5;\{n_{ij}\})\bigg|_{p_5=s_{14}=0} \nonumber \\
    &+\sin{(\pi(s_{13}+a))}\mathcal{I}^{\text{op}}_5(2,3,4,1,5;\{n_{ij}\})\bigg|_{p_5=s_{14}=0}\bigg). \label{B-B}
\end{align}
The above expression is evaluated at $s_{14}=0$ and $p_5=0$ providing that $s_{12}=s_{24}$, $s_{13}=s_{34}$ if all the strings are in the same level of string mass spectrum. The set $\{n_{ij}\}$ is $\{ n_{12}=-a, n_{13}=-b, n_{14}=a+b \}$.

Besides, we recognize a similarity between the expression of our correction terms (\ref{e63}) 
and the infinite tube amplitudes presented in \cite{Stieberger:2016lng} albeit an infinite sum. For more points of string amplitudes, we would also obtain correction terms similar to the four-point scenario but with exponentially more complicated expressions as one needs to deal with multiple contours of integration together with multiple breach points and cuts in the bulk of the complex plane. Nevertheless, this issue can be overlooked if we consider the field theory limit, $\alpha' \to 0$, to which it relates Einstein's theory with Einstein-Yang-Mills theory. In this limit, the correction terms in (\ref{additional term2}) are at the order of $\alpha'$ assuming that the open string amplitudes are at the leading order. Also according to \cite{STIEBERGER2016104}, the mixed string integral $\mathcal{F}_n$ is at the same order as the open string amplitudes. This means that we can ignore the correction terms in this limit. This argument persists to higher-point string amplitudes as well. This is because there always appears contribution from contours of integration along infinite tubes which are there to avoid crossing the branch cuts inside the complex plane. Such contours would provide a factor of $\sin(\pi s_{ij})$ every time the contours encircling around the branch cuts. This results in suppression when taking the field theory limit.

\section{Conclusions}\label{sec13}
Using analytic continuation of complex variables, we successfully factorize closed string amplitudes into those involving ($n-2$) open strings and a single closed string. Our derivation begins with a detailed examination of four-point closed string amplitudes and subsequently extends to generalize the results for arbitrary $n$-point amplitudes. The expressions for four, five, and six-point amplitudes are presented in (\ref{e9}), (\ref{e20}), and (\ref{relation 6pt}), respectively, while the relations for $n$ strings are expressed in (\ref{e30}).

In addition, our findings reveal the presence of correction terms stemming from integration along infinite tubes in the complex plane. These corrections are introduced when considering the existence of branch points in the bulk of the complex plane during the Wick rotation. The correction terms for the four-point case are explicitly given in (\ref{additional term2}). To capture these correction terms, we define the function $\mathcal{B}(x,y)$, which can be expressed in terms of integral forms for open string amplitudes, as shown in (\ref{B-B}).

Importantly, in the field theory limit, the correction terms become negligible, being at least of the order of $\alpha'$. This justifies the validity of the established connections in low-energy limits and enhances the applicability of our results.

In summary, our work achieves the factorization of closed string amplitudes, provides explicit derivations for specific cases, addresses correction terms arising from complex variable transformations, and justifies the neglect of these corrections in the low-energy limit. 

\appendix
\section{Polarizations and kinematic factors} \label{A1}
As mentioned in Section \ref{sec1}, the function $F_n(z_i,\bar{z}_i)$ contains the polarizations and kinematic factors of the external closed string states. During the factorization, the closed string polarizations which are encoded in the function $F_n$ are decomposed to those open and closed strings in the mixed amplitudes captured by $K_n$ and $\widetilde{K}_n$.

For tachyons, $F_n=K_n=\widetilde{K}_n=1$. In case of the first excited states, $F_n(z_i,\bar{z}_i)$ is given by 
\begin{align}
    F_n(z_i,\bar{z}_i) &= \text{exp} \Bigg{\{}\sum_{i>j}\frac{\zeta_i \cdot \zeta_j}{(z_i-z_j)^2}-\sqrt{\alpha'}\sum_{i\neq j}\frac{k_i \cdot \zeta_j}{(z_i-z_j)} \nonumber \\ 
    &+\sum_{i>j}\frac{\tilde{\zeta}_i \cdot \tilde{\zeta}_j}{(\bar{z}_i-\bar{z}_j)^2} -\sqrt{\alpha'}\sum_{i\neq j}\frac{k_i \cdot \tilde{\zeta}_j}{(\bar{z}_i-\bar{z}_j)}\Bigg{\}}\Bigg{|}_{\text{linear in }\zeta_i,\tilde{\zeta}_i}, \label{eA1}
\end{align}
where $\zeta_i \otimes \tilde{\zeta}_i$ = $\chi_i$  are polarization of external closed string states. For a scattering including $n-2$ open strings with polarizations $\zeta_i$, for $i\in \{1,\ldots, n-2\}$, and one closed string with a polarization $\chi=\vartheta\otimes \tilde\vartheta$, the function $K_n(x_i)$ reads
\begin{align}
    &\text{exp} \Bigg{\{}  \sum_{i>j}^{n-2}\frac{\zeta_i \cdot \zeta_j}{(x_i-x_j)^2}-\sqrt{\alpha'}\sum_{i\neq j}^{n-2} \frac{k_i \cdot \zeta_j}{(x_i-x_j)}+\sum_{j=1}^{n-3}\Bigg{[} \frac{\vartheta\cdot \zeta_j}{(x_j-z)^2}+\frac{\tilde\vartheta\cdot \zeta_j}{(x_j-\bar{z})^2} \nonumber \\
     &-\sqrt{\alpha'}\Bigg{(}\frac{k_{n-1}\cdot \zeta_j}{(\bar{z}-\eta_j)} +\frac{k_{n-2}\cdot \zeta_j}{(z-\eta_j)} +\frac{k_j \cdot \vartheta}{(x_j-z)} +\frac{k_j \cdot \tilde\vartheta}{(x_j-\bar{z})}\Bigg{)}\Bigg{]} \nonumber \\
     &+\frac{\vartheta \cdot \tilde\vartheta}{(z-\bar{z})^2}-\sqrt{\alpha'}\Bigg{(}\frac{k_{n-1}\cdot \vartheta}{(\bar{z}-z)} +\frac{k_{n-2}\cdot \tilde\vartheta}{(z-\bar{z})} \Bigg{)} \Bigg{\}}\Bigg{|}_{\text{linear in } \zeta_i,\vartheta,\tilde\vartheta } \label{mix K}
\end{align}
where the closed string momentum is $k_{n-2}+k_{n-1}$.

We can factorize (\ref{eA1}) into a product of two functions, i.e. $F_n(z_i,\bar{z}_i)$ = $f_n(z_i) \times \tilde{f}_n(\bar{z})$. Then, we follow the similar calculation done in section \ref{sec2} utilizing the analytic continuation of complex variables and defining new variables as in (\ref{new var2}), we obtain 

\begin{align}
    f_n(\eta_i) = &\text{exp} \Bigg{\{} \sum_{i>j}^{n-3}\frac{\zeta_i \cdot \zeta_j}{(\eta_i-\eta_j)^2}-\sqrt{\alpha'}\sum_{i\neq j}^{n-3} \frac{k_i \cdot \zeta_j}{(\eta_i-\eta_j)}+\sum_{j=1}^{n-3}\Bigg{[} \frac{\zeta_{n-1}\cdot \zeta_j}{(\eta_j+i)^2}+\frac{\zeta_{n-2}\cdot \zeta_j}{(\eta_j-i)^2} \nonumber \\
     &-\sqrt{\alpha'}\Bigg{(}\frac{k_{n-1}\cdot \zeta_j}{(-i-\eta_j)} +\frac{k_{n-2}\cdot \zeta_j}{(i-\eta_j)} +\frac{k_j \cdot \zeta_{n-1}}{(\eta_j+i)} +\frac{k_j \cdot \zeta_{n-2}}{(\eta_j-i)}\Bigg{)}\Bigg{]} \nonumber \\
     &+\frac{\zeta_{n-1} \cdot \zeta_{n-2}}{(-2i)^2}-\sqrt{\alpha'}\Bigg{(}\frac{k_{n-1}\cdot \zeta_{n-2}}{(-2i)} +\frac{k_{n-2}\cdot \zeta_{n-1}}{(2i)} \Bigg{)} \Bigg{\}}\Bigg{|}_{\text{linear in } \zeta_i} \label{eA4}
\end{align}
and
\begin{align}
    \tilde{f}_n(\xi_i) = &\text{exp} \Bigg{\{} \sum_{i>j}^{n-3}\frac{\tilde{\zeta}_i \cdot \tilde{\zeta}_j}{(\xi_i-\xi_j)^2}-\sqrt{\alpha'}\sum_{i\neq j}^{n-3} \frac{k_i \cdot \tilde{\zeta}_j}{(\xi_i-\xi_j)}+\sum_{j=1}^{n-3}\Bigg{[} \frac{\tilde{\zeta}_{n-1}\cdot \tilde{\zeta}_j}{(\xi_j-i)^2}+\frac{\tilde{\zeta}_{n-2}\cdot \tilde{\zeta}_j}{(\xi_j+i)^2} \nonumber \\
     &-\sqrt{\alpha'}\Bigg{(}\frac{k_{n-1}\cdot \tilde{\zeta}_j}{(i-\xi_j)} +\frac{k_{n-2}\cdot \tilde{\zeta}_j}{(-i-\xi_j)} +\frac{k_j \cdot \tilde{\zeta}_{n-1}}{(\xi_j-i)} +\frac{k_j \cdot \tilde{\zeta}_{n-2}}{(\xi_j+i)}\Bigg{)}\Bigg{]} \nonumber \\
     &+\frac{\tilde{\zeta}_{n-1} \cdot \tilde{\zeta}_{n-2}}{(2i)^2}- \sqrt{\alpha'}\Bigg{(}\frac{k_{n-1}\cdot \tilde{\zeta}_{n-2}}{(2i)} +\frac{k_{n-2}\cdot \tilde{\zeta}_{n-1}}{(-2i)} \Bigg{)} \Bigg{\}}\Bigg{|}_{\text{linear in } \tilde{\zeta}_i}.\label{eA5}
\end{align}
Note that the points $z_{n-2}, z_{n-1}$ and $z_n$ are fixed to $i, -i$ and $\infty$ respectively. It is not hard to see that the functions $f_n$ and $\tilde{f}_n$ can be identified as $K_n$ and $\tilde{K}_n$ (\ref{mix K}) where now the closed string polartization regarding $K_n$ and $\tilde{K}_n$ are $\zeta_{n-1}\otimes \zeta_{n-2}$ and $\tilde{\zeta}_{n-1}\otimes \tilde{\zeta}_{n-2}$ respectively in mixed string amplitudes. 



\bibliography{sn-bibliography}

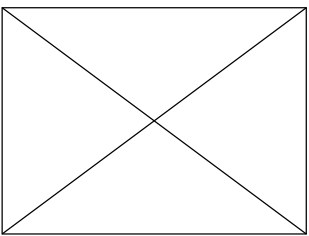
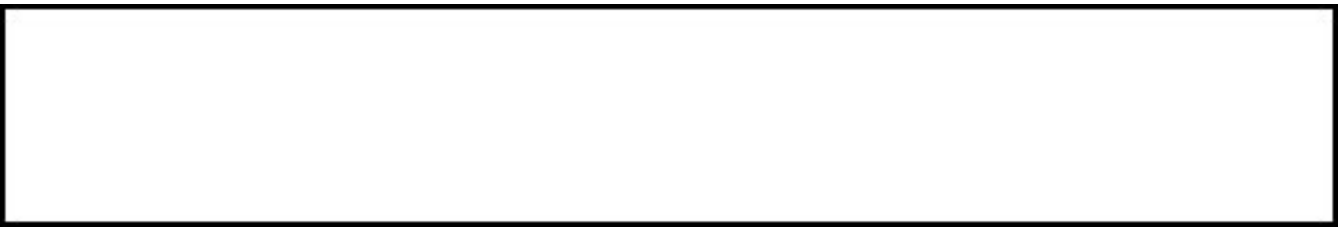
\end{document}